\documentclass[sigconf]{acmart}
\newcommand{\modelname}{\textsf{GraphAU}\xspace}
\AtBeginDocument{%
  \providecommand\BibTeX{{%
    \normalfont B\kern-0.5em{\scshape i\kern-0.25em b}\kern-0.8em\TeX}}}

\usepackage{multirow}
\usepackage{enumitem}
\usepackage{xspace}
\usepackage{color}
\usepackage{amsmath}

\copyrightyear{2023}
\acmYear{2023}
\setcopyright{acmlicensed}\acmConference[CIKM '23]{Proceedings of the 32nd ACM International Conference on Information and Knowledge Management}{October 21--25, 2023}{Birmingham, United Kingdom}
\acmBooktitle{Proceedings of the 32nd ACM International Conference on Information and Knowledge Management (CIKM '23), October 21--25, 2023, Birmingham, United Kingdom}
\acmPrice{15.00}
\acmDOI{10.1145/3583780.3615185}
\acmISBN{979-8-4007-0124-5/23/10}

\begin{document}

\title{Graph-based Alignment and Uniformity for Recommendation}
\author{Liangwei Yang}
\affiliation{%
  \institution{University of Illinois Chicago}
  \city{Chicago}
  \country{USA}}
\email{lyang84@uic.edu}

\author{Zhiwei~Liu}
\affiliation{%
  \institution{Salesforce AI Research}
  \city{Palo Alto}
  \country{USA}
}
\email{zhiweiliu@salesforce.com}

\author{Chen~Wang}
\affiliation{%
  \institution{University of Illinois at Chicago}
  \city{Chicago}
  \country{USA}}
\email{cwang266@uic.edu}

\author{Mingdai~Yang}
\email{myang72@uic.edu}
\author{Xiaolong~Liu}
\email{xliu262@uic.edu}
\affiliation{%
  \institution{University of Illinois at Chicago}
  \city{Chicago}
  \country{USA}}

\author{Jing Ma}
\affiliation{%
  \institution{University of Electronic Science and Technology of China}
  \city{Chengdu}
  \country{China}}
\email{jingma@uestc.edu.cn}

\author{Philip S. Yu}
\affiliation{%
  \institution{University of Illinois Chicago}
  \city{Chicago}
  \country{USA}}
\email{psyu@cs.uic.edu}

\renewcommand{\shortauthors}{Liangwei Yang et al.}

\begin{abstract}
Collaborative filtering-based recommender systems (RecSys) rely on learning representations for users and items to predict preferences accurately. Representation learning on the hypersphere is a promising approach due to its desirable properties, such as alignment and uniformity. However, the sparsity issue arises when it encounters RecSys. To address this issue, we propose a novel approach, graph-based alignment and uniformity (\modelname), that explicitly considers high-order connectivities in the user-item bipartite graph. \modelname aligns the user/item embedding to the dense vector representations of high-order neighbors using a neighborhood aggregator, eliminating the need to compute the burdensome alignment to high-order neighborhoods individually. To address the discrepancy in alignment losses, \modelname includes a layer-wise alignment pooling module to integrate alignment losses layer-wise. Experiments on four datasets show that \modelname significantly alleviates the sparsity issue and achieves state-of-the-art performance. We open-source GraphAU at \textcolor{blue}{\url{https://github.com/YangLiangwei/GraphAU}}.
\end{abstract}




\maketitle

\section{Introduction}
\begin{figure}
    \begin{center}
    \includegraphics[width=.3\textwidth]{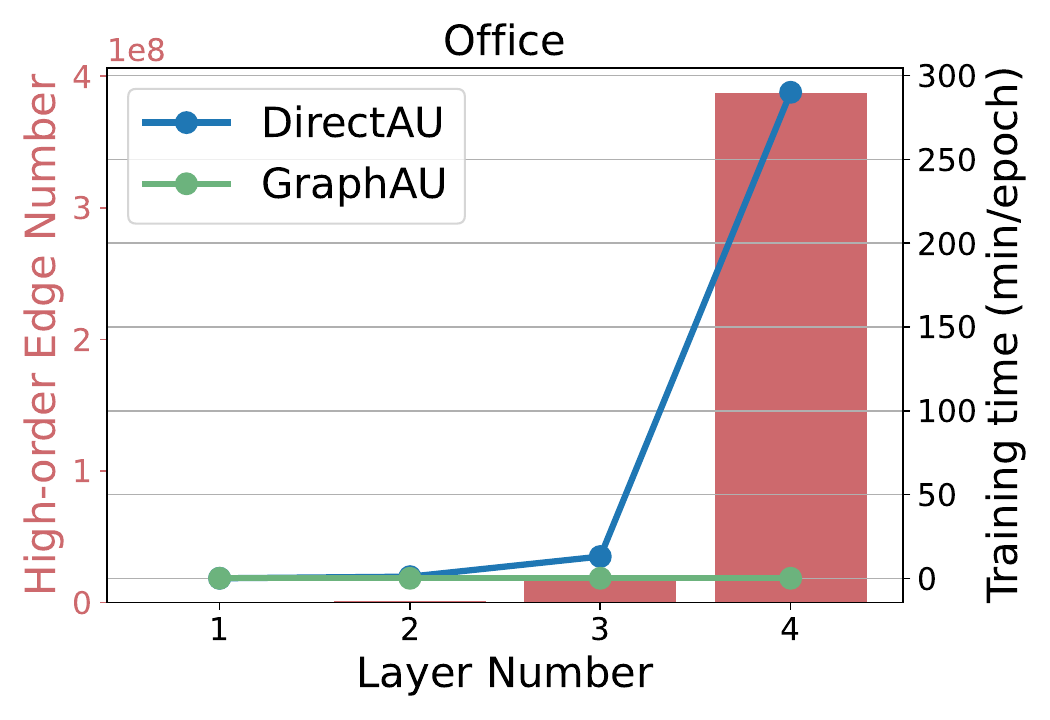}
    \end{center}
    \caption{The scalability problem of DirectAU, and the efficiency of \modelname. GraphAU nearly keeps constant Training time (line plot) with High-order Edge Number (Bar Plot).}
    \label{fig:scale}
\end{figure}
In regard to the overwhelming online information~\cite{mayer2013big},
recommender systems (RecSys) assist users effortlessly discovering items that align with their interests~\cite{lu2012recommender,wu2022feedrec,yang2022large,gu2020hierarchical}. 
The core of most existing RecSys is learning representations for users and items.
By embedding past user-item interactions into dense vectors~\cite{yang2022dgrec}, RecSys create sophisticated representations of users and items that capture the nuances of their preferences and characteristics.
The effectiveness of learning user/item representations is heavily influenced by the choice of loss functions~\cite{li2022autolossgen,mfbpr,yang2021consisrec,cao2007learning,tang2021multisample,tang2023ranking}. 
Recent studies explain that the desirable performance of contrastive loss results from the alignment of positive pairs and uniformity of data~\cite{wang2020understanding,pu2022alignment,dufumier2021conditional}.
This observation motivates the development of an innovative technique, Direct Alignment and Uniformity (DirectAU)~\cite{wang2022towards} in the field of RecSys. 
DirectAU employs alignment loss to 
improve the normalized element-wise similarity 
between a user's representation and those of the items he/she has interacted with.
It also adopts uniformity loss to ensure that the user and item representations are evenly distributed and inherently distinguishable from each other.
\begin{figure*}[!hbt]
    \centering
    \includegraphics[width=0.9\linewidth]{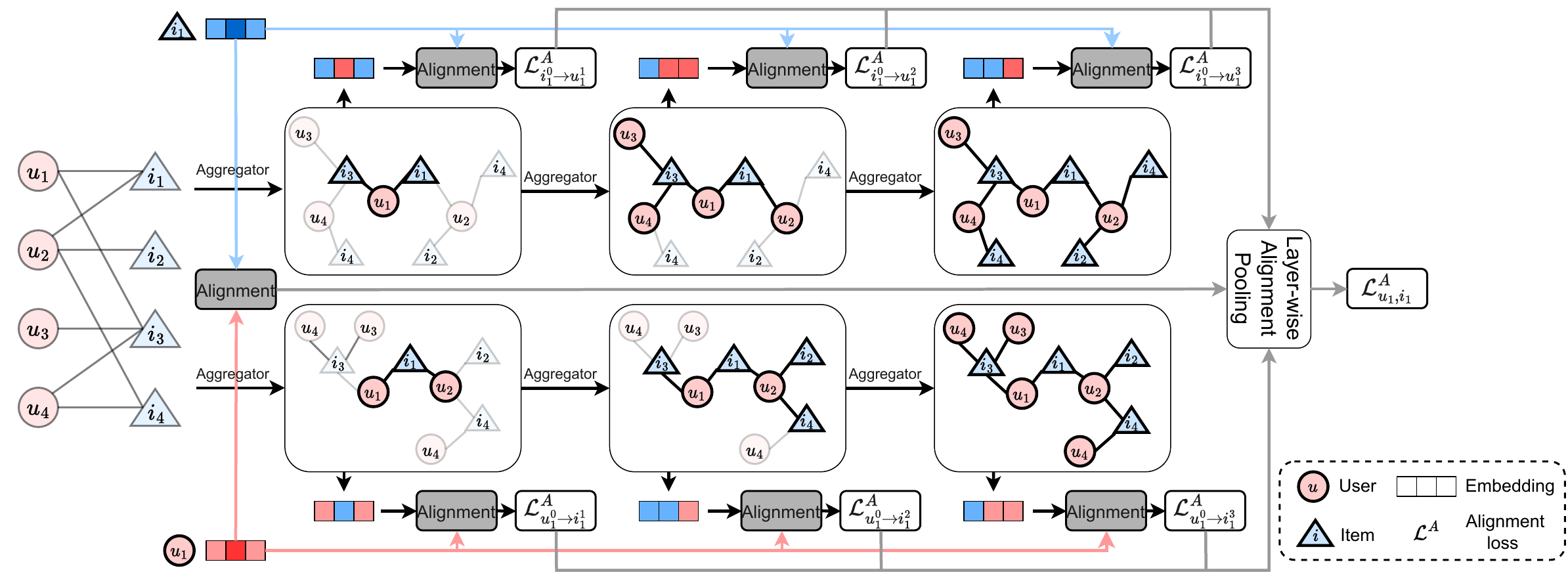}
    \caption{An example of computing alignment loss between $u_1$ and $i_1$ considering neighborhoods within $3$ hops. \modelname applies $3$ Aggregators on the user-item bipartite graph to obtain a dense neighborhoods representation of $u_1$/$i_1$ within hops from $1$ to $3$. The output of each Aggregator is explicitly aligned with $u_1$ or $i_1$ correspondingly before the pooling module.}
    \label{framework}
\end{figure*}

Despite the effectiveness, we contend that directly utilizing alignment between user-item pairs, DirectAU neglects the critical data sparsity~\cite{guo2017resolving} issue in RecSys. 
The alignment signals between users and his/her direct interacted items are rather sparse, which hinders the effectiveness of alignment objective. 
Inspired by existing works~\cite{wang2021graph,he2020lightgcn,wang2019neural,chen2020revisiting,wang2020disentangled,shen2021powerful,zhang2022geometric,yang2023ranking,liu2021kg} that leverage high-order connections in user-item bipartite graph, we propose to devise a novel \modelname to enhance current alignment loss.
The direct alignment of high-order connectivities encounters two primary challenges. Firstly, the scalability of high-order connections poses a significant hurdle. 
As shown in Figure~\ref{fig:scale}, the number of high-order edges increases exponentially with more hops, and the training time of DirectAU also increases exponentially. 
Consequently, direct alignment of all high-order edges can be impractical and time-consuming. 
Secondly, there is a discrepancy in different orders. Alignment losses from different hop neighborhoods exhibit neighborhood similarity and influence scope discrepancies. Low-order neighborhoods in the user-item bipartite graph are typically more pertinent to the center node. Besides, the influence of alignment loss from low-order neighbors is only propagated to a small portion of nodes compared with high-order neighbors.
It is necessary to consider the discrepancies of alignment losses in different orders.

This paper proposes a novel approach, called graph-based alignment and uniformity (\modelname), to address the sparsity issue by explicitly considering high-order connectivities in the user-item bipartite graph. To overcome the scalability issue, \modelname aligns the user/item embedding to the dense vector representations of high-order neighbors instead of directly aligning to high-order neighborhoods individually. To achieve this, several layers of aggregators are used to obtain the dense representation of neighborhoods within different hops. Then, the user/item embedding is directly aligned to the dense representation of high-order neighborhoods of the connected item/user. This approach eliminates the need for individually computing the burdensome alignment to high-order neighborhoods and resolves the scalability issue through neighborhood aggregation. As shown in Figure~\ref{fig:scale}, the training time per epoch of \modelname nearly keeps constant with the considered exponential increased high-order edges. Compared with DirectAU, \modelname greatly reduces the training time and enables the alignment toward high-order neighborhoods.
To address the discrepancy in alignment losses, \modelname includes a layer-wise alignment pooling module to integrate alignment losses layer-wise. A modification factor $\alpha$ is introduced to adjust the weight of alignment loss from different layers. \modelname significantly alleviates the sparsity issue and achieves state-of-the-art performance on four datasets with varying scales. The contributions of our paper are summarized as follows:
\begin{itemize}[leftmargin=*]
    \item We propose a novel approach, named \modelname, that explicitly considers high-order connectivities in the user-item bipartite graph to address the sparsity issue.
    \item To address the discrepancy in alignment losses, we propose a simple and effective layer-wise alignment pooling module to integrate alignment losses layer-wise.
    \item We conduct extensive experiments on $4$ real-world datasets with varying scales and demonstrate the effectiveness of \modelname.
\end{itemize}
\section{Preliminaries}
\textbf{Problem Statement.}
In the context of a recommendation task, the objective is to generate a list of top $k$ items that a given user $u$ is likely to be interested in. This is based on the historical interactions between users and items, which are represented by an interaction matrix $\textbf{R}$ of size $\left| \mathcal{U} \right| \times \left| \mathcal{I} \right|$. The set of users is denoted by $\mathcal{U} = \{u_1,u_2,...,u_{\left | \mathcal{U} \right|}\}$ and the set of items by $\mathcal{I} = \{i_1,i_2,...,i_{\left | \mathcal{I} \right|}\}$. 
In this paper, we only consider the implicit feedback, and the interaction matrix $\textbf{R}$ is binary, where $R_{u,i}=1$ if user $u$ has interacted with item $i$, and $R_{u,i}=0$ otherwise.
To model the historical interactions, a user-item bipartite graph $\mathcal{G}=(\mathcal{V}, \mathcal{E})$ is constructed, where $\mathcal{V}= \mathcal{U}\cup \mathcal{I}$ and there is an edge $(u,i) \in \mathcal{E}$ between $u$ and $i$ if $R_{u,i}=1$. The objective of the recommender system is to learn from this graph $\mathcal{G}$ and generate a ranked list of potential items for each user $u$.

\textbf{Alignment and Uniformity in RecSys.}
The two properties~\cite{wang2020understanding,gao2021simcse} are introduced into RecSys by DirectAU~\cite{wang2022towards} that considers the user-item interactions as positive pairs. They are optimized by two different losses. Alignment loss is calculated by:
\begin{equation}\label{eq:alignment}
    \mathcal{L}^{A}=\frac{1}{|\mathcal{E}|}\sum_{(u,i)\in \mathcal{E}}\mathcal{L}^{A}_{u, i}=
    \frac{1}{|\mathcal{E}|}\sum_{(u,i)\in \mathcal{E}}||\mathbf{e}_u - \mathbf{e}_i||^2,
\end{equation}
where $\mathcal{L}^{A}_{u, i}$ is the alignment loss from $u$ to $i$, and $\mathbf{e}_u$/$\mathbf{e}_i$ is user/item representation. Alignment aims to strengthen the normalized element-wise similarity between the user's and his/her interacted items' representation. Uniformity loss is calculated as $    \mathcal{L}^{U}=\frac{1}{2}(\mathcal{L}^{U}_{\mathcal{U}}+\mathcal{L}^{U}_{\mathcal{I}})$,
where $\mathcal{L}^{U}_{\mathcal{U}}$ is the uniformity loss for user set $\mathcal{U}$, denoted as:
\begin{equation}
    \mathcal{L}^{U}_{\mathcal{U}} = \log \frac{1}{|\mathcal{U}|^2}\sum_{u\in\mathcal{U}}\sum_{u^*\in \mathcal{U}}e^{-2||e_u-e_u^*||}.
\end{equation}
Item set uniformity $\mathcal{L}^{U}_{\mathcal{I}}$ is calculated analogously.
Uniformity targets distributing user/item representation uniformly and distinguishable.
It is utilized as a regularization to avoid a trivial solution (\textit{i.e.} all same) of users/item embedding through alignment loss.

\begin{table*}[htbp]
\caption{Overall comparison, the best and second-best results are in bold and underlined, respectively}
\label{tab:comparison}
\small
\begin{tabular}{lcccccccccccc}
\toprule

\multirow{2}{*}{Method} & \multicolumn{3}{c}{Office} & \multicolumn{3}{c}{Toys} & \multicolumn{3}{c}{Beauty} & \multicolumn{3}{c}{Gowalla} \\

\cmidrule(r){2-4} \cmidrule(r){5-7} \cmidrule(r){8-10} \cmidrule{11-13}
& R@20 & HR@20 & N@20 & R@20 & HR@20 & N@20 & R@20 & HR@20 & N@20 & R@20 & HR@20 & N@20 \\
\midrule
MF-BPR & 0.0818 & \underline{0.1762} & 0.0441  & 0.0755 & 0.1193 & 0.0393 & 0.0580 & 0.0936 & 0.0303 & 0.0824 & 0.3213 & 0.0627 \\
NGCF  & 0.0753 & 0.1689 & 0.0403 & 0.0668 & 0.1048 & 0.0346 & 0.0843 & 0.1327 & 0.0453 & 0.0856 & 0.3288 & 0.0648 \\
LightGCN  & 0.0711 & 0.1554 & 0.0380 & 0.0937 & 0.1483 & 0.0490 & 0.1035 & 0.1628 & 0.0546 & 0.0720 & 0.2901 & 0.0552 \\
UltraGCN &  0.0796 & 0.1639 & 0.0439 & 0.0928 & 0.1481 & 0.0493 & 0.1043 & 0.1619 & 0.0560 & 0.0887 & 0.3411 & 0.0668 \\
\hline
DirectAU & \underline{0.0839} & 0.1682 & \underline{0.0474} & \underline{0.0985} & \underline{0.1559} & \underline{0.0531} & \underline{0.1074} & \underline{0.1677} & \underline{0.0583} & \underline{0.1134} & \underline{0.4051} & \underline{0.0847} \\

GraphAU & \textbf{0.0979} & \textbf{0.2003} & \textbf{0.0539} & \textbf{0.1041} & \textbf{0.1637} & \textbf{0.0551} & \textbf{0.1124} & \textbf{0.1752} & \textbf{0.0599} & \textbf{0.1174} & \textbf{0.4136} & \textbf{0.0855} \\
\hline
Improvement & 16.64\% & 13.68\% & 13.71\% & 5.71\% & 5.04\% & 3.81\% & 4.63\% & 4.47\% & 2.78\% & 3.52\% & 2.11\% & 1.01\% \\
\bottomrule 
\end{tabular} 
\end{table*}

\section{Proposed Model}
The framework of \modelname is shown in Figure~\ref{framework}. Two designed modules are described as follows.
\subsection{Multi-hop Neighborhood Alignment}
The proposed Multi-hop Neighborhood Alignment Module addresses the scalability issue of aligning high-order neighborhoods. This is achieved through an aggregator to obtain dense vector representations of neighborhoods within multiple hops. By doing so, the scalability issue is addressed. The module directly computes the alignment loss towards the high-order dense neighborhood representations, eliminating the need to align high-order neighborhoods, which can be burdensome individually.

Similar to the approach used in learning representations of words and graphs~\cite{vaswani2017attention,liu2018multi}, embedding techniques have been widely adopted in recommender systems~\cite{mfbpr,he2020lightgcn,liu2021dense}. This involves the use of an embedding layer, which serves as a look-up table to map user and item IDs to dense vectors, denoted as $\mathbf{E}^{(0)}=\left(\mathbf{e}_1^{(0)}, \mathbf{e}_2^{(0)}, \ldots, \mathbf{e}_{|\mathcal{U}|+|\mathcal{I}|}^{(0)}\right)$,
 where $\mathbf{e}^{(0)}\in\mathbb{R}^d$ is a $d$-dimensional dense vector corresponding to a specific user or item. The embedding is then fed into an aggregator for information aggregation. The resulting output from the embedding layer is commonly referred to as the $0$-th layer output, denoted as $\mathbf{e}_i^{(0)}$.
 We then compute the dense vector representation of neighborhoods via an aggregator as:
 \begin{align}
    \mathbf{e}_{u}^{(l+1)} = \text{Aggregator}^{(l+1)} (\{ \mathbf{e}_{i}^{(l)} \mid i \in \mathcal{N}_u \}),
\end{align}
where $\mathbf{e}_{u}^{(l)}$ indicates node $u$'s embedding on the $l$-th layer, $\mathcal{N}_u$ is the neighbor set of node $u$, $\mbox{Aggregator}^{(l)}(\cdot)$ aggregates neighbors' embeddings into a single vector for layer $l$. The aggregator for items is computed similarly. The aggregator can be any pooling function over the neighborhood's representation to a dense vector as long as it is non-parametric. 
Each aggregation would generate one embedding vector for each user/item node. Embedding generated from different layers is the dense representation of neighbors within different hops. After $L$-th layer convolution, we can have the multi-hop neighborhood alignment loss as Equation~\ref{eq:alignment} from different layers, as illustrated in Figure~\ref{framework}:
\begin{equation}
\begin{aligned}
    (\mathcal{L}^{A}_{u^{(0)}, i^{(1)}},\mathcal{L}^{A}_{u^{(0)}, i^{(2)}},...,\mathcal{L}^{A}_{u^{(0)}, i^{(L)}}), \\
    (\mathcal{L}^{A}_{i^{(0)}, u^{(1)}}, \mathcal{L}^{A}_{i^{(0)}, u^{(2)}},..., \mathcal{L}^{A}_{i^{(0)}, u^{(L)}}),
\end{aligned}
\end{equation}
where $\mathcal{L}^{A}_{u^{(0)}, i^{(L)}}=||\mathbf{e}_u^{(0)} - \mathbf{e}_i^{(L)}||^2$ is the alignment loss from $u$'s embedding $\mathbf{e}_u^{(0)}$ to the $i$'s representation after $L$ layers aggregator $\mathbf{e}_i^{(L)}$, which is the dense representation of neighbors within $L$ hops. \modelname aligns the embedding layer directly to high-order neighborhood representations, facilitating the embedding layer's awareness of the high-order information of the counterpart.

\subsection{Layer-wise Alignment Pooling}
The alignment losses obtained from different layers vary regarding their relevance to the center node and the scope of their influence. To address this, a layer-wise alignment pooling module is proposed to integrate the discrepant losses. Firstly, the user and item alignment losses from the same layer are combined through averaging: $\mathcal{L}_{u,i}^{A,(L)}=\frac{1}{2}(\mathcal{L}_{u^{(0)},i^{(L)}}^{A} + \mathcal{L}_{i^{(0)},u^{(L)}}^{A})$.
We then use a modification factor $\alpha$ to adjust the alignment weight for each layer and combine them with a weighted sum:
\begin{equation}
    \mathcal{L}^{A}_{u,i}=\mathcal{L}_{u,i}^{A,(0)} + \alpha\mathcal{L}_{u,i}^{A,(1)} +  \dots + \alpha^{L}\mathcal{L}_{u,i}^{A,(L)},
\end{equation}
where $\mathcal{L}_{u,i}^{A,(0)}=||\mathbf{e}_u^{(0)} - \mathbf{e}_i^{(0)}||^2$ is the direct alignment loss. \modelname incorporates a weight factor, denoted by $\alpha$, to balance the importance of relevant and high-order alignment losses. Specifically, when $\alpha<1$, the method emphasizes the relevant alignment loss and gradually assigns less weight to high-order alignment loss. When $\alpha=1$, no discrepancy is made for different layers. Finally, when $\alpha>1$, the method prioritizes the influence scope and assigns more weight to the high-order alignment loss.

A uniformity loss is included to prevent all the embedding from aligning identically and enable easier distinction between users and items. The final loss function is as follows:
\begin{equation}
    \mathcal{L} = \frac{1}{|\mathcal{E}|}\sum_{(u,i)\in \mathcal{E}}\mathcal{L}^{A}_{u, i} + \frac{\gamma}{2} \cdot (\mathcal{L}^{U}_{\mathcal{U}} + \mathcal{L}^{U}_{\mathcal{I}}),
\end{equation}
where $\gamma$ is a trade-off hyper-parameter for alignment and uniformity. After optimization, the rating score from $u$ to $i$ is directly computed as their embedding dot product as $R_{u,i}={\mathbf{e}_u^{(0)}}^{\top}\mathbf{e}_{i}^{(0)}$.


\section{Experiments}

\subsection{Experimental Setup}
\textbf{Datasets}: We conduct experiments on $4$ real-world datasets with varying scales, including Amazon Office, Toys, and Beauty datasets~\cite{mcauley2015image,he2016ups} and one check-in dataset Gowalla~\cite{cho2011friendship}. We randomly split the dataset into training set ($60\%$), validation set ($20\%$), and test set ($20\%$). We select
Recall@20 (R@20), Hit Ratio@20 (HR@20) and nDCG@20 (N@20) as metrics. 
\textbf{Baselines}: To justify the effectiveness of \modelname, we compare it with $5$ baseline methods. MF-BPR~\cite{mfbpr}, NGCF~\cite{wang2019neural}, LightGCN~\cite{he2020lightgcn}, UltraGCN~\cite{mao2021ultragcn} and DirectAU~\cite{wang2022towards}. To make a fair comparison, we fix the embedding size as $32$, and tune all the methods based on the same grid search. The learning rate is searched in $\{0.1, 0.05, 0.01, 0.005\}$. We use Adam optimizer~\cite{kingma2014adam} and search weight decay in $\{0.0, 1e^{-2}, 1e^{-4}, 1e^{-6}, 1e^{-8}\}$. For NGCF, LightGCN, and \modelname, layer number is tuned from $1$ to $4$. For DirectAU and \modelname, $\gamma$ is tuned from $0.0$ to $1.0$ with a step of $0.1$. For \modelname, we use Light Graph Convolution~\cite{he2020lightgcn} as the aggregator and tune $\alpha$ from $0.0$ to $2.0$ with a step of $0.1$. We apply early stop with $10$ epochs to avoid over-fitting.
\subsection{Performance Evaluation}
Experiment results are shown in Table~\ref{tab:comparison}. We can have the following observations. 
1) \modelname achieves the best performance on all the datasets. On the Office dataset, \modelname surpasses DirectAU for $16.64\%$ in R@20. It justifies the advantages of considering high-order alignment in RecSys.
 2) DirectAU performs better than other graph-based baselines. It shows a direct utilization of alignment and uniformity on user-item pairs can learn an informative user/item embedding, which justifies the advantages of alignment and uniformity-based user/item representation learning in RecSys. 
3) The performance of graph-based baselines varies with different datasets. For example, NGCF overpasses LightGCN on Office and Gowalla datasets while failing on Toys and Beauty dataset. It shows pure graph-based methods are sensitive to datasets.
\subsection{Model Analysis}
\begin{figure}[t]
    \begin{center}
    \includegraphics[width=.2\textwidth]{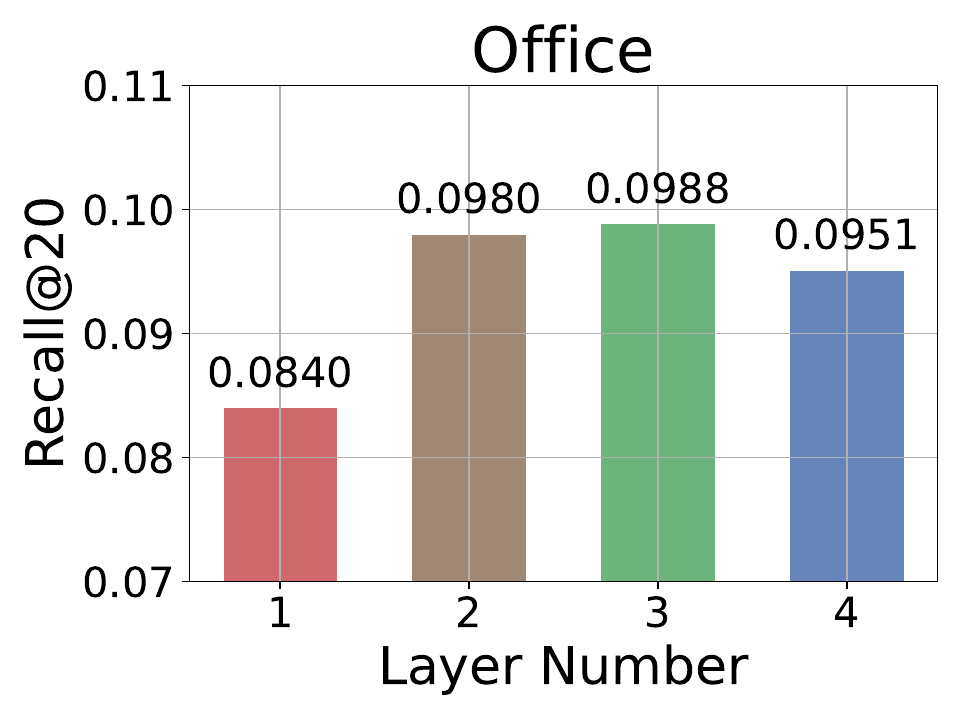}
    \includegraphics[width=.2\textwidth]{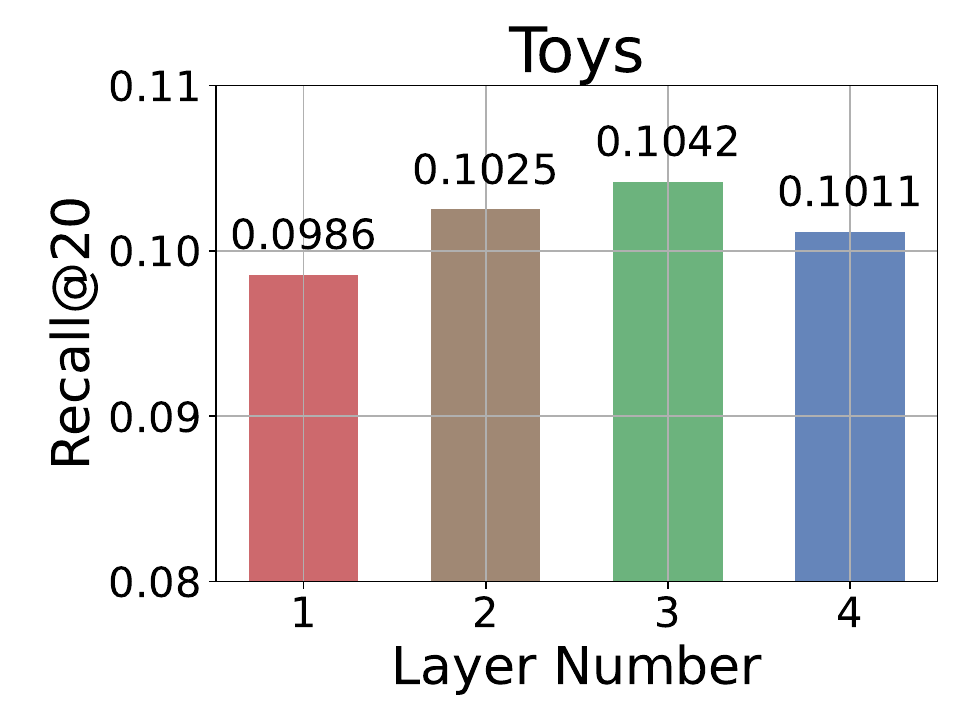}
    \includegraphics[width=.2\textwidth]{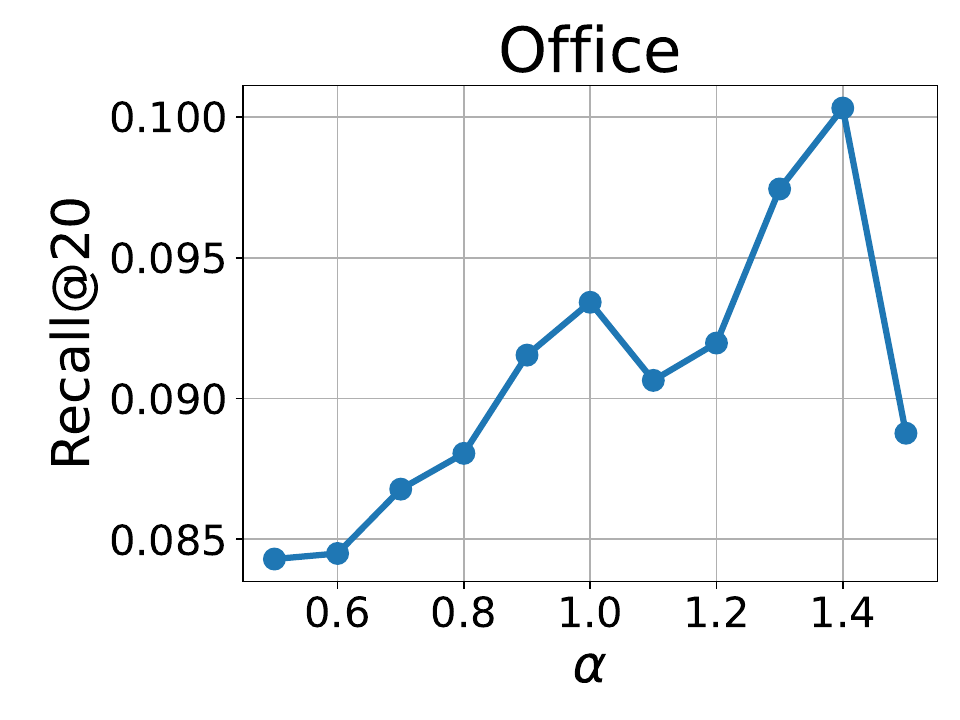}
    \includegraphics[width=.2\textwidth]{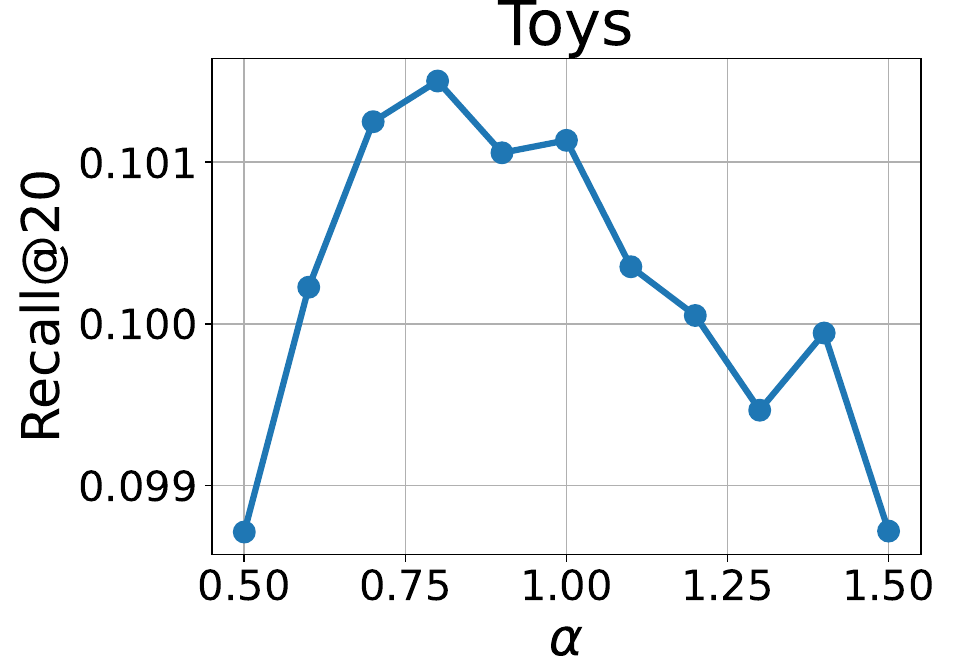}
    \includegraphics[width=.2\textwidth]{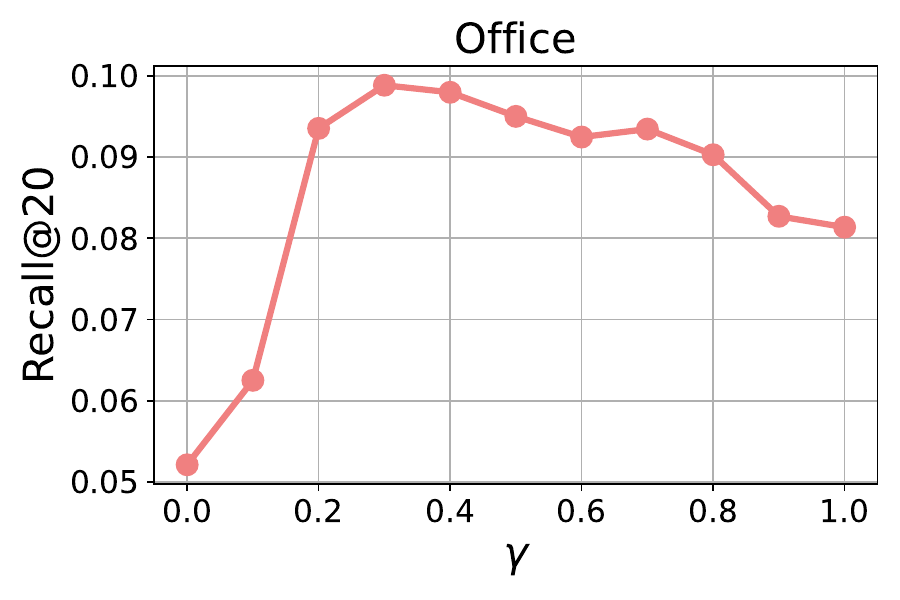}
    \includegraphics[width=.2\textwidth]{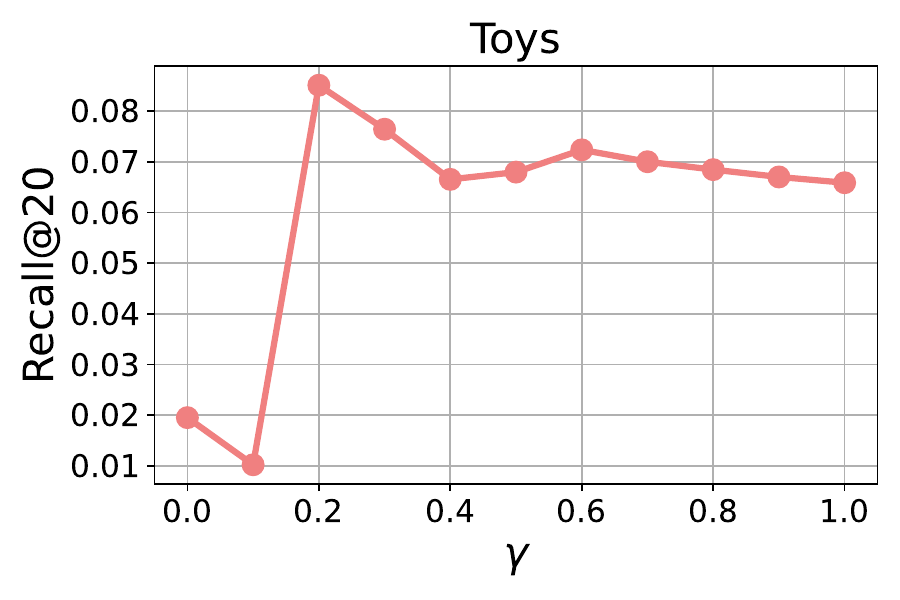}
    \end{center}
    \caption{Parameter sensitivity of layer number, $\alpha$ and $\gamma$}
    \label{fig:analysis}
\end{figure}
We further analyze the influential hyper-parameters layer number $L$, modification factor $\alpha$, and alignment/uniformity trade-off $\gamma$ of \modelname. Experiment results are shown in Figure~\ref{fig:analysis}. We can have the following observations.
1) $L$: With the increase of layer number, the performance of \modelname first increases to its peak and then decreases on both datasets. The initial increase shows \modelname alleviates the sparsity issue by aligning high-order neighborhoods' representation, while the following decrease indicates excessive high-order alignment leads to the performance drop. 2) $\alpha$: 
When we conduct experiments on $\alpha$, we fix the number of layers as $4$ to observe its influence better. Experiments show $\alpha$ impacts the performance of \modelname greatly. An interesting finding is that the peak point of $\alpha$ can be smaller (as in Toys) or larger (as in Office) than $1$ based on different datasets. It shows for different datasets, \modelname adapts on different aspects. When $\alpha < 1$, \modelname assigns decaying weights for high-order alignment to focus more on local relevant neighborhoods. When $\alpha>1$, \modelname increases the weights for high-order alignment, which shows it focuses more on the influence scope of alignment loss. 3) $\gamma$: It is the most influential hyper-parameter on all datasets, which requires careful tuning for \modelname.
Similar to high-order alignment in \modelname, we also investigate the impact of high-order uniformity by adding uniformity loss on the aggregated user/item high-order embedding. Results are shown in Table~\ref{tab:uniformity}. We can see that adding high-order uniformity loss deteriorates the performance of \modelname. It is because the graph explicitly represents only the alignment information between users and items. Enforcing excessive regularization of high-order uniformity is not a favorable option for \modelname.
\begin{table}[t]
  \caption{High-order uniformity experiment on Office.}
  \label{tab:uniformity}
  \small
  \begin{tabular}{l| c c c}
        \toprule
        Method & R@20 & HR@20 & N@20 \\
        \hline
        GraphAU & \textbf{0.0979}  & \textbf{0.2003}  & \textbf{0.0539}  \\
        + $1$-st order uniformity & 0.0849  & 0.1704 & 0.0482 \\
        + $2$-nd order uniformity & 0.0857  & 0.1732  & 0.0473 \\
        + $3$-rd order uniformity & 0.0799  & 0.1600   & 0.0440  \\
        \bottomrule
  \end{tabular}
\end{table}
\section{Conclusion}
In this paper, we identify, study, and cope with the sparsity issue when alignment/uniformity encounters RecSys. We propose a solution called \modelname, which focuses on explicitly aligning user/item embeddings while taking into account high-order connectivities in the user-item bipartite graph. 
To address scalability, \modelname aligns towards the dense vector neighborhood representation obtained by an aggregator, rather than aligning with high-order neighborhoods individually. 
We also introduce a modification factor that effectively integrates discrepant alignment losses from different layers. 
Our experiments demonstrate the benefits of aligning high-order neighborhoods and the effectiveness of \modelname.
\begin{acks}
This work is supported in part by NSF under grants III-1763325, III-1909323,  III-2106758, and SaTC-1930941.
\end{acks}

\bibliographystyle{ACM-Reference-Format}
\bibliography{sample-base}


\begin{thebibliography}{35}


\ifx \showCODEN    \undefined \def \showCODEN     #1{\unskip}     \fi
\ifx \showDOI      \undefined \def \showDOI       #1{#1}\fi
\ifx \showISBNx    \undefined \def \showISBNx     #1{\unskip}     \fi
\ifx \showISBNxiii \undefined \def \showISBNxiii  #1{\unskip}     \fi
\ifx \showISSN     \undefined \def \showISSN      #1{\unskip}     \fi
\ifx \showLCCN     \undefined \def \showLCCN      #1{\unskip}     \fi
\ifx \shownote     \undefined \def \shownote      #1{#1}          \fi
\ifx \showarticletitle \undefined \def \showarticletitle #1{#1}   \fi
\ifx \showURL      \undefined \def \showURL       {\relax}        \fi
\providecommand\bibfield[2]{#2}
\providecommand\bibinfo[2]{#2}
\providecommand\natexlab[1]{#1}
\providecommand\showeprint[2][]{arXiv:#2}

\bibitem[Cao et~al\mbox{.}(2007)]%
        {cao2007learning}
\bibfield{author}{\bibinfo{person}{Zhe Cao}, \bibinfo{person}{Tao Qin},
  \bibinfo{person}{Tie-Yan Liu}, \bibinfo{person}{Ming-Feng Tsai}, {and}
  \bibinfo{person}{Hang Li}.} \bibinfo{year}{2007}\natexlab{}.
\newblock \showarticletitle{Learning to rank: from pairwise approach to
  listwise approach}. In \bibinfo{booktitle}{\emph{Proceedings of the 24th
  international conference on Machine learning}}. \bibinfo{pages}{129--136}.
\newblock


\bibitem[Chen et~al\mbox{.}(2020)]%
        {chen2020revisiting}
\bibfield{author}{\bibinfo{person}{Lei Chen}, \bibinfo{person}{Le Wu},
  \bibinfo{person}{Richang Hong}, \bibinfo{person}{Kun Zhang}, {and}
  \bibinfo{person}{Meng Wang}.} \bibinfo{year}{2020}\natexlab{}.
\newblock \showarticletitle{Revisiting graph based collaborative filtering: A
  linear residual graph convolutional network approach}. In
  \bibinfo{booktitle}{\emph{Proceedings of the AAAI conference on artificial
  intelligence}}, Vol.~\bibinfo{volume}{34}. \bibinfo{pages}{27--34}.
\newblock


\bibitem[Cho et~al\mbox{.}(2011)]%
        {cho2011friendship}
\bibfield{author}{\bibinfo{person}{Eunjoon Cho}, \bibinfo{person}{Seth~A
  Myers}, {and} \bibinfo{person}{Jure Leskovec}.}
  \bibinfo{year}{2011}\natexlab{}.
\newblock \showarticletitle{Friendship and mobility: user movement in
  location-based social networks}. In \bibinfo{booktitle}{\emph{Proceedings of
  the 17th ACM SIGKDD international conference on Knowledge discovery and data
  mining}}. \bibinfo{pages}{1082--1090}.
\newblock


\bibitem[Dufumier et~al\mbox{.}(2021)]%
        {dufumier2021conditional}
\bibfield{author}{\bibinfo{person}{Benoit Dufumier}, \bibinfo{person}{Pietro
  Gori}, \bibinfo{person}{Julie Victor}, \bibinfo{person}{Antoine Grigis},
  {and} \bibinfo{person}{Edouard Duchesnay}.} \bibinfo{year}{2021}\natexlab{}.
\newblock \showarticletitle{Conditional Alignment and Uniformity for
  Contrastive Learning with Continuous Proxy Labels}.
\newblock \bibinfo{journal}{\emph{arXiv preprint arXiv:2111.05643}}
  (\bibinfo{year}{2021}).
\newblock


\bibitem[Gao et~al\mbox{.}(2021)]%
        {gao2021simcse}
\bibfield{author}{\bibinfo{person}{Tianyu Gao}, \bibinfo{person}{Xingcheng
  Yao}, {and} \bibinfo{person}{Danqi Chen}.} \bibinfo{year}{2021}\natexlab{}.
\newblock \showarticletitle{Simcse: Simple contrastive learning of sentence
  embeddings}.
\newblock \bibinfo{journal}{\emph{arXiv preprint arXiv:2104.08821}}
  (\bibinfo{year}{2021}).
\newblock


\bibitem[Gu et~al\mbox{.}(2020)]%
        {gu2020hierarchical}
\bibfield{author}{\bibinfo{person}{Yulong Gu}, \bibinfo{person}{Zhuoye Ding},
  \bibinfo{person}{Shuaiqiang Wang}, {and} \bibinfo{person}{Dawei Yin}.}
  \bibinfo{year}{2020}\natexlab{}.
\newblock \showarticletitle{Hierarchical user profiling for e-commerce
  recommender systems}. In \bibinfo{booktitle}{\emph{Proceedings of the 13th
  International Conference on Web Search and Data Mining}}.
  \bibinfo{pages}{223--231}.
\newblock


\bibitem[Guo et~al\mbox{.}(2017)]%
        {guo2017resolving}
\bibfield{author}{\bibinfo{person}{Guibing Guo}, \bibinfo{person}{Huihuai Qiu},
  \bibinfo{person}{Zhenhua Tan}, \bibinfo{person}{Yuan Liu},
  \bibinfo{person}{Jing Ma}, {and} \bibinfo{person}{Xingwei Wang}.}
  \bibinfo{year}{2017}\natexlab{}.
\newblock \showarticletitle{Resolving data sparsity by multi-type auxiliary
  implicit feedback for recommender systems}.
\newblock \bibinfo{journal}{\emph{Knowledge-Based Systems}}
  \bibinfo{volume}{138} (\bibinfo{year}{2017}), \bibinfo{pages}{202--207}.
\newblock


\bibitem[He and McAuley(2016)]%
        {he2016ups}
\bibfield{author}{\bibinfo{person}{Ruining He} {and} \bibinfo{person}{Julian
  McAuley}.} \bibinfo{year}{2016}\natexlab{}.
\newblock \showarticletitle{Ups and downs: Modeling the visual evolution of
  fashion trends with one-class collaborative filtering}. In
  \bibinfo{booktitle}{\emph{proceedings of the 25th international conference on
  world wide web}}. \bibinfo{pages}{507--517}.
\newblock


\bibitem[He et~al\mbox{.}(2020)]%
        {he2020lightgcn}
\bibfield{author}{\bibinfo{person}{Xiangnan He}, \bibinfo{person}{Kuan Deng},
  \bibinfo{person}{Xiang Wang}, \bibinfo{person}{Yan Li},
  \bibinfo{person}{Yongdong Zhang}, {and} \bibinfo{person}{Meng Wang}.}
  \bibinfo{year}{2020}\natexlab{}.
\newblock \showarticletitle{Lightgcn: Simplifying and powering graph
  convolution network for recommendation}. In
  \bibinfo{booktitle}{\emph{Proceedings of the 43rd International ACM SIGIR
  conference on research and development in Information Retrieval}}.
  \bibinfo{pages}{639--648}.
\newblock


\bibitem[Kingma and Ba(2014)]%
        {kingma2014adam}
\bibfield{author}{\bibinfo{person}{Diederik~P Kingma} {and}
  \bibinfo{person}{Jimmy Ba}.} \bibinfo{year}{2014}\natexlab{}.
\newblock \showarticletitle{Adam: A method for stochastic optimization}.
\newblock \bibinfo{journal}{\emph{arXiv preprint arXiv:1412.6980}}
  (\bibinfo{year}{2014}).
\newblock


\bibitem[Li et~al\mbox{.}(2022)]%
        {li2022autolossgen}
\bibfield{author}{\bibinfo{person}{Zelong Li}, \bibinfo{person}{Jianchao Ji},
  \bibinfo{person}{Yingqiang Ge}, {and} \bibinfo{person}{Yongfeng Zhang}.}
  \bibinfo{year}{2022}\natexlab{}.
\newblock \showarticletitle{AutoLossGen: Automatic Loss Function Generation for
  Recommender Systems}. In \bibinfo{booktitle}{\emph{Proceedings of the 45th
  International ACM SIGIR Conference on Research and Development in Information
  Retrieval}}. \bibinfo{pages}{1304--1315}.
\newblock


\bibitem[Liu et~al\mbox{.}(2021a)]%
        {liu2021dense}
\bibfield{author}{\bibinfo{person}{Ye Liu}, \bibinfo{person}{Kazuma Hashimoto},
  \bibinfo{person}{Yingbo Zhou}, \bibinfo{person}{Semih Yavuz},
  \bibinfo{person}{Caiming Xiong}, {and} \bibinfo{person}{Philip~S Yu}.}
  \bibinfo{year}{2021}\natexlab{a}.
\newblock \showarticletitle{Dense hierarchical retrieval for open-domain
  question answering}.
\newblock \bibinfo{journal}{\emph{arXiv preprint arXiv:2110.15439}}
  (\bibinfo{year}{2021}).
\newblock


\bibitem[Liu et~al\mbox{.}(2018)]%
        {liu2018multi}
\bibfield{author}{\bibinfo{person}{Ye Liu}, \bibinfo{person}{Lifang He},
  \bibinfo{person}{Bokai Cao}, \bibinfo{person}{Philip Yu},
  \bibinfo{person}{Ann Ragin}, {and} \bibinfo{person}{Alex Leow}.}
  \bibinfo{year}{2018}\natexlab{}.
\newblock \showarticletitle{Multi-view multi-graph embedding for brain network
  clustering analysis}. In \bibinfo{booktitle}{\emph{Proceedings of the AAAI
  conference on artificial intelligence}}, Vol.~\bibinfo{volume}{32}.
\newblock


\bibitem[Liu et~al\mbox{.}(2021b)]%
        {liu2021kg}
\bibfield{author}{\bibinfo{person}{Ye Liu}, \bibinfo{person}{Yao Wan},
  \bibinfo{person}{Lifang He}, \bibinfo{person}{Hao Peng}, {and}
  \bibinfo{person}{S~Yu Philip}.} \bibinfo{year}{2021}\natexlab{b}.
\newblock \showarticletitle{Kg-bart: Knowledge graph-augmented bart for
  generative commonsense reasoning}. In \bibinfo{booktitle}{\emph{Proceedings
  of the AAAI Conference on Artificial Intelligence}},
  Vol.~\bibinfo{volume}{35}. \bibinfo{pages}{6418--6425}.
\newblock


\bibitem[L{\"u} et~al\mbox{.}(2012)]%
        {lu2012recommender}
\bibfield{author}{\bibinfo{person}{Linyuan L{\"u}},
  \bibinfo{person}{Mat{\'u}{\v{s}} Medo}, \bibinfo{person}{Chi~Ho Yeung},
  \bibinfo{person}{Yi-Cheng Zhang}, \bibinfo{person}{Zi-Ke Zhang}, {and}
  \bibinfo{person}{Tao Zhou}.} \bibinfo{year}{2012}\natexlab{}.
\newblock \showarticletitle{Recommender systems}.
\newblock \bibinfo{journal}{\emph{Physics reports}} \bibinfo{volume}{519},
  \bibinfo{number}{1} (\bibinfo{year}{2012}), \bibinfo{pages}{1--49}.
\newblock


\bibitem[Mao et~al\mbox{.}(2021)]%
        {mao2021ultragcn}
\bibfield{author}{\bibinfo{person}{Kelong Mao}, \bibinfo{person}{Jieming Zhu},
  \bibinfo{person}{Xi Xiao}, \bibinfo{person}{Biao Lu},
  \bibinfo{person}{Zhaowei Wang}, {and} \bibinfo{person}{Xiuqiang He}.}
  \bibinfo{year}{2021}\natexlab{}.
\newblock \showarticletitle{UltraGCN: ultra simplification of graph
  convolutional networks for recommendation}. In
  \bibinfo{booktitle}{\emph{Proceedings of the 30th ACM International
  Conference on Information \& Knowledge Management}}.
  \bibinfo{pages}{1253--1262}.
\newblock


\bibitem[Mayer-Sch{\"o}nberger and Cukier(2013)]%
        {mayer2013big}
\bibfield{author}{\bibinfo{person}{Viktor Mayer-Sch{\"o}nberger} {and}
  \bibinfo{person}{Kenneth Cukier}.} \bibinfo{year}{2013}\natexlab{}.
\newblock \bibinfo{booktitle}{\emph{Big data: A revolution that will transform
  how we live, work, and think}}.
\newblock \bibinfo{publisher}{Houghton Mifflin Harcourt}.
\newblock


\bibitem[McAuley et~al\mbox{.}(2015)]%
        {mcauley2015image}
\bibfield{author}{\bibinfo{person}{Julian McAuley},
  \bibinfo{person}{Christopher Targett}, \bibinfo{person}{Qinfeng Shi}, {and}
  \bibinfo{person}{Anton Van Den~Hengel}.} \bibinfo{year}{2015}\natexlab{}.
\newblock \showarticletitle{Image-based recommendations on styles and
  substitutes}. In \bibinfo{booktitle}{\emph{Proceedings of the 38th
  international ACM SIGIR conference on research and development in information
  retrieval}}. \bibinfo{pages}{43--52}.
\newblock


\bibitem[Pu et~al\mbox{.}(2022)]%
        {pu2022alignment}
\bibfield{author}{\bibinfo{person}{Shi Pu}, \bibinfo{person}{Kaili Zhao}, {and}
  \bibinfo{person}{Mao Zheng}.} \bibinfo{year}{2022}\natexlab{}.
\newblock \showarticletitle{Alignment-Uniformity aware Representation Learning
  for Zero-shot Video Classification}. In \bibinfo{booktitle}{\emph{Proceedings
  of the IEEE/CVF Conference on Computer Vision and Pattern Recognition}}.
  \bibinfo{pages}{19968--19977}.
\newblock


\bibitem[Rendle et~al\mbox{.}(2012)]%
        {mfbpr}
\bibfield{author}{\bibinfo{person}{Steffen Rendle}, \bibinfo{person}{Christoph
  Freudenthaler}, \bibinfo{person}{Zeno Gantner}, {and} \bibinfo{person}{Lars
  Schmidt-Thieme}.} \bibinfo{year}{2012}\natexlab{}.
\newblock \showarticletitle{BPR: Bayesian personalized ranking from implicit
  feedback}.
\newblock \bibinfo{journal}{\emph{arXiv preprint arXiv:1205.2618}}
  (\bibinfo{year}{2012}).
\newblock


\bibitem[Shen et~al\mbox{.}(2021)]%
        {shen2021powerful}
\bibfield{author}{\bibinfo{person}{Yifei Shen}, \bibinfo{person}{Yongji Wu},
  \bibinfo{person}{Yao Zhang}, \bibinfo{person}{Caihua Shan},
  \bibinfo{person}{Jun Zhang}, \bibinfo{person}{B~Khaled Letaief}, {and}
  \bibinfo{person}{Dongsheng Li}.} \bibinfo{year}{2021}\natexlab{}.
\newblock \showarticletitle{How powerful is graph convolution for
  recommendation?}. In \bibinfo{booktitle}{\emph{Proceedings of the 30th ACM
  international conference on information \& knowledge management}}.
  \bibinfo{pages}{1619--1629}.
\newblock


\bibitem[Tang et~al\mbox{.}(2023)]%
        {tang2023ranking}
\bibfield{author}{\bibinfo{person}{Hao Tang}, \bibinfo{person}{Guoshuai Zhao},
  \bibinfo{person}{Yujiao He}, \bibinfo{person}{Yuxia Wu}, {and}
  \bibinfo{person}{Xueming Qian}.} \bibinfo{year}{2023}\natexlab{}.
\newblock \showarticletitle{Ranking-based contrastive loss for recommendation
  systems}.
\newblock \bibinfo{journal}{\emph{Knowledge-Based Systems}}
  \bibinfo{volume}{261} (\bibinfo{year}{2023}), \bibinfo{pages}{110180}.
\newblock


\bibitem[Tang et~al\mbox{.}(2021)]%
        {tang2021multisample}
\bibfield{author}{\bibinfo{person}{Hao Tang}, \bibinfo{person}{Guoshuai Zhao},
  \bibinfo{person}{Yuxia Wu}, {and} \bibinfo{person}{Xueming Qian}.}
  \bibinfo{year}{2021}\natexlab{}.
\newblock \showarticletitle{Multisample-based contrastive loss for top-k
  recommendation}.
\newblock \bibinfo{journal}{\emph{IEEE Transactions on Multimedia}}
  (\bibinfo{year}{2021}).
\newblock


\bibitem[Vaswani et~al\mbox{.}(2017)]%
        {vaswani2017attention}
\bibfield{author}{\bibinfo{person}{Ashish Vaswani}, \bibinfo{person}{Noam
  Shazeer}, \bibinfo{person}{Niki Parmar}, \bibinfo{person}{Jakob Uszkoreit},
  \bibinfo{person}{Llion Jones}, \bibinfo{person}{Aidan~N Gomez},
  \bibinfo{person}{{\L}ukasz Kaiser}, {and} \bibinfo{person}{Illia
  Polosukhin}.} \bibinfo{year}{2017}\natexlab{}.
\newblock \showarticletitle{Attention is all you need}.
\newblock \bibinfo{journal}{\emph{Advances in neural information processing
  systems}}  \bibinfo{volume}{30} (\bibinfo{year}{2017}).
\newblock


\bibitem[Wang et~al\mbox{.}(2022)]%
        {wang2022towards}
\bibfield{author}{\bibinfo{person}{Chenyang Wang}, \bibinfo{person}{Yuanqing
  Yu}, \bibinfo{person}{Weizhi Ma}, \bibinfo{person}{Min Zhang},
  \bibinfo{person}{Chong Chen}, \bibinfo{person}{Yiqun Liu}, {and}
  \bibinfo{person}{Shaoping Ma}.} \bibinfo{year}{2022}\natexlab{}.
\newblock \showarticletitle{Towards Representation Alignment and Uniformity in
  Collaborative Filtering}. In \bibinfo{booktitle}{\emph{Proceedings of the
  28th ACM SIGKDD Conference on Knowledge Discovery and Data Mining}}.
  \bibinfo{pages}{1816--1825}.
\newblock


\bibitem[Wang et~al\mbox{.}(2021)]%
        {wang2021graph}
\bibfield{author}{\bibinfo{person}{Shoujin Wang}, \bibinfo{person}{Liang Hu},
  \bibinfo{person}{Yan Wang}, \bibinfo{person}{Xiangnan He},
  \bibinfo{person}{Quan~Z Sheng}, \bibinfo{person}{Mehmet~A Orgun},
  \bibinfo{person}{Longbing Cao}, \bibinfo{person}{Francesco Ricci}, {and}
  \bibinfo{person}{Philip~S Yu}.} \bibinfo{year}{2021}\natexlab{}.
\newblock \showarticletitle{Graph learning based recommender systems: A
  review}.
\newblock \bibinfo{journal}{\emph{arXiv preprint arXiv:2105.06339}}
  (\bibinfo{year}{2021}).
\newblock


\bibitem[Wang and Isola(2020)]%
        {wang2020understanding}
\bibfield{author}{\bibinfo{person}{Tongzhou Wang} {and}
  \bibinfo{person}{Phillip Isola}.} \bibinfo{year}{2020}\natexlab{}.
\newblock \showarticletitle{Understanding contrastive representation learning
  through alignment and uniformity on the hypersphere}. In
  \bibinfo{booktitle}{\emph{International Conference on Machine Learning}}.
  PMLR, \bibinfo{pages}{9929--9939}.
\newblock


\bibitem[Wang et~al\mbox{.}(2019)]%
        {wang2019neural}
\bibfield{author}{\bibinfo{person}{Xiang Wang}, \bibinfo{person}{Xiangnan He},
  \bibinfo{person}{Meng Wang}, \bibinfo{person}{Fuli Feng}, {and}
  \bibinfo{person}{Tat-Seng Chua}.} \bibinfo{year}{2019}\natexlab{}.
\newblock \showarticletitle{Neural graph collaborative filtering}. In
  \bibinfo{booktitle}{\emph{Proceedings of the 42nd international ACM SIGIR
  conference on Research and development in Information Retrieval}}.
  \bibinfo{pages}{165--174}.
\newblock


\bibitem[Wang et~al\mbox{.}(2020)]%
        {wang2020disentangled}
\bibfield{author}{\bibinfo{person}{Xiang Wang}, \bibinfo{person}{Hongye Jin},
  \bibinfo{person}{An Zhang}, \bibinfo{person}{Xiangnan He},
  \bibinfo{person}{Tong Xu}, {and} \bibinfo{person}{Tat-Seng Chua}.}
  \bibinfo{year}{2020}\natexlab{}.
\newblock \showarticletitle{Disentangled graph collaborative filtering}. In
  \bibinfo{booktitle}{\emph{Proceedings of the 43rd international ACM SIGIR
  conference on research and development in information retrieval}}.
  \bibinfo{pages}{1001--1010}.
\newblock


\bibitem[Wu et~al\mbox{.}(2022)]%
        {wu2022feedrec}
\bibfield{author}{\bibinfo{person}{Chuhan Wu}, \bibinfo{person}{Fangzhao Wu},
  \bibinfo{person}{Tao Qi}, \bibinfo{person}{Qi Liu}, \bibinfo{person}{Xuan
  Tian}, \bibinfo{person}{Jie Li}, \bibinfo{person}{Wei He},
  \bibinfo{person}{Yongfeng Huang}, {and} \bibinfo{person}{Xing Xie}.}
  \bibinfo{year}{2022}\natexlab{}.
\newblock \showarticletitle{Feedrec: News feed recommendation with various user
  feedbacks}. In \bibinfo{booktitle}{\emph{Proceedings of the ACM Web
  Conference 2022}}. \bibinfo{pages}{2088--2097}.
\newblock


\bibitem[Yang et~al\mbox{.}(2021)]%
        {yang2021consisrec}
\bibfield{author}{\bibinfo{person}{Liangwei Yang}, \bibinfo{person}{Zhiwei
  Liu}, \bibinfo{person}{Yingtong Dou}, \bibinfo{person}{Jing Ma}, {and}
  \bibinfo{person}{Philip~S Yu}.} \bibinfo{year}{2021}\natexlab{}.
\newblock \showarticletitle{Consisrec: Enhancing gnn for social recommendation
  via consistent neighbor aggregation}. In
  \bibinfo{booktitle}{\emph{Proceedings of the 44th international ACM SIGIR
  conference on Research and development in information retrieval}}.
  \bibinfo{pages}{2141--2145}.
\newblock


\bibitem[Yang et~al\mbox{.}(2022a)]%
        {yang2022large}
\bibfield{author}{\bibinfo{person}{Liangwei Yang}, \bibinfo{person}{Zhiwei
  Liu}, \bibinfo{person}{Yu Wang}, \bibinfo{person}{Chen Wang},
  \bibinfo{person}{Ziwei Fan}, {and} \bibinfo{person}{Philip~S Yu}.}
  \bibinfo{year}{2022}\natexlab{a}.
\newblock \showarticletitle{Large-scale personalized video game recommendation
  via social-aware contextualized graph neural network}. In
  \bibinfo{booktitle}{\emph{Proceedings of the ACM Web Conference 2022}}.
  \bibinfo{pages}{3376--3386}.
\newblock


\bibitem[Yang et~al\mbox{.}(2022b)]%
        {yang2022dgrec}
\bibfield{author}{\bibinfo{person}{Liangwei Yang}, \bibinfo{person}{Shengjie
  Wang}, \bibinfo{person}{Yunzhe Tao}, \bibinfo{person}{Jiankai Sun},
  \bibinfo{person}{Xiaolong Liu}, \bibinfo{person}{Philip~S Yu}, {and}
  \bibinfo{person}{Taiqing Wang}.} \bibinfo{year}{2022}\natexlab{b}.
\newblock \showarticletitle{DGRec: Graph Neural Network for Recommendation with
  Diversified Embedding Generation}.
\newblock \bibinfo{journal}{\emph{arXiv preprint arXiv:2211.10486}}
  (\bibinfo{year}{2022}).
\newblock


\bibitem[Yang et~al\mbox{.}(2023)]%
        {yang2023ranking}
\bibfield{author}{\bibinfo{person}{Mingdai Yang}, \bibinfo{person}{Zhiwei Liu},
  \bibinfo{person}{Liangwei Yang}, \bibinfo{person}{Xiaolong Liu},
  \bibinfo{person}{Chen Wang}, \bibinfo{person}{Hao Peng}, {and}
  \bibinfo{person}{Philip~S Yu}.} \bibinfo{year}{2023}\natexlab{}.
\newblock \showarticletitle{Ranking-based group identification via factorized
  attention on social tripartite graph}. In
  \bibinfo{booktitle}{\emph{Proceedings of the Sixteenth ACM International
  Conference on Web Search and Data Mining}}. \bibinfo{pages}{769--777}.
\newblock


\bibitem[Zhang et~al\mbox{.}(2022)]%
        {zhang2022geometric}
\bibfield{author}{\bibinfo{person}{Yiding Zhang}, \bibinfo{person}{Chaozhuo
  Li}, \bibinfo{person}{Xing Xie}, \bibinfo{person}{Xiao Wang},
  \bibinfo{person}{Chuan Shi}, \bibinfo{person}{Yuming Liu},
  \bibinfo{person}{Hao Sun}, \bibinfo{person}{Liangjie Zhang},
  \bibinfo{person}{Weiwei Deng}, {and} \bibinfo{person}{Qi Zhang}.}
  \bibinfo{year}{2022}\natexlab{}.
\newblock \showarticletitle{Geometric Disentangled Collaborative Filtering}. In
  \bibinfo{booktitle}{\emph{Proceedings of the 45th International ACM SIGIR
  Conference on Research and Development in Information Retrieval}}.
  \bibinfo{pages}{80--90}.
\newblock


\end{thebibliography}

\appendix

\end{document}